\documentclass[num-refs, twocolumn]{wiley-article}

\usepackage{siunitx}
\usepackage{fixltx2e}

\papertype{Original Article}
\paperfield{Journal Section}

\title{An Ultra-Compact Single FeFET Binary and Multi-Bit Associative Search Engine}

\abbrevs{FeFET, Ferroelectric field effect transistor; CAM, content addressable memory; HDC, hyperdimensional computing; ReRAM, resistive random access memory; HD, hamming distance.}

\author[1]{Xunzhao Yin}
\author[2]{Franz M\"uller}
\author[1]{Qingrong Huang}
\author[2]{Chao Li}
\author[3]{Mohsen Imani}
\author[1]{Zeyu Yang}
\author[1]{Jiahao Cai}
\author[2]{Maximilian Lederer}
\author[2]{Ricardo Olivo}
\author[2]{Nellie Laleni}
\author[4]{Shan Deng}
\author[4]{Zijian Zhao}
\author[1]{Cheng Zhuo}
\author[2]{Thomas K\"ampfe}
\author[4]{Kai Ni}

\affil[1]{Zhejiang University, Hangzhou, Zhejiang, China}
\affil[2]{Fraunhofer IPMS, Dresden, Germany}
\affil[3]{University of California Irvine, Irvine, CA, USA}
\affil[4]{Rochester Institute of Technology, Rochester, NY, USA}

\corraddress{Cheng Zhuo, Thomas K\"ampfe, Kai Ni}
\corremail{czhuo@zju.edu.cn, thomas.kaempfe@ipms.fraunhofer.de, kai.ni@rit.edu}

\runningauthor{Xunzhao Yin et al.}

\begin{document}

\begin{frontmatter}
\maketitle

\begin{abstract}
Content addressable memory (CAM) is widely used in associative search tasks for its highly parallel pattern matching capability. To accommodate the increasingly complex and data-intensive pattern matching tasks, it is critical to keep improving the CAM density to enhance the performance and area efficiency. 
In this work, we demonstrate: i) a novel ultra-compact 1FeFET CAM design that enables parallel associative search and in-memory hamming distance calculation; ii) a multi-bit CAM for exact search using the same CAM cell; iii) compact device designs that integrate the series resistor current limiter into the intrinsic FeFET structure to turn the 1FeFET1R into an effective 1FeFET cell; iv) a successful 2-step search operation and a sufficient sensing margin of the proposed binary and multi-bit 1FeFET1R CAM array with sizes of practical interests in both experiments and simulations, given the existing unoptimized FeFET device variation; v) 89.9x speedup and 66.5x energy efficiency improvement over the state-of-the-art alignment tools on GPU in accelerating genome pattern matching applications through the hyperdimensional computing paradigm.

\end{abstract}
\end{frontmatter}

\section{Introduction}
\label{sec:introduction}

In the era of artificial intelligence (AI) and Internet of Things (IoT), the ever growing amount of data generated  by various machine learning (ML) models and devices in edges and data centers has placed severe demands on efficient computational hardware and architectures to support high performance applications.
However, the conventional Von Neumann architectures are consuming significant energy costs and latency due to the massive data transfer between storage and processing units, which is the so-called memory wall issues.
Emerging computational accelerators, specifically in-memory computing (IMC) circuits and architectures, have been intensively proposed and studied to address the memory wall issues by limiting the data movements and replacing sequential operations with parallel data analytic operations, thus improving the performance and energy efficiency of the computational cores \cite{ielmini2018memory, verma2019memory, sebastian2020memory}. 
Moreover, with the utilization of emerging non-volatile memory (NVM) devices, e.g., resistive random access memory (ReRAM), spin torque transfer magnetic random access memory (STT-MRAM) and ferroelectric field effect transistor (FeFET), IMC solutions have been further improved in terms of area, information density and energy efficiency over traditional complementary metal oxide semiconductor (CMOS) based cores \cite{ielmini2018memory, sebastian2020memory}. 
Various IMC elements based on NVMs have been studied, especially crossbar structures which have been used to accelerate the core matrix multiplication operations in data-intensive tasks, e.g., neural networks, signal processing and differential equation solving, etc. \cite{ielmini2018memory, sebastian2020memory}.

\begin{figure*}[h]
	\centering
	\vspace{-1ex}
	\includegraphics [width=1.0\linewidth]{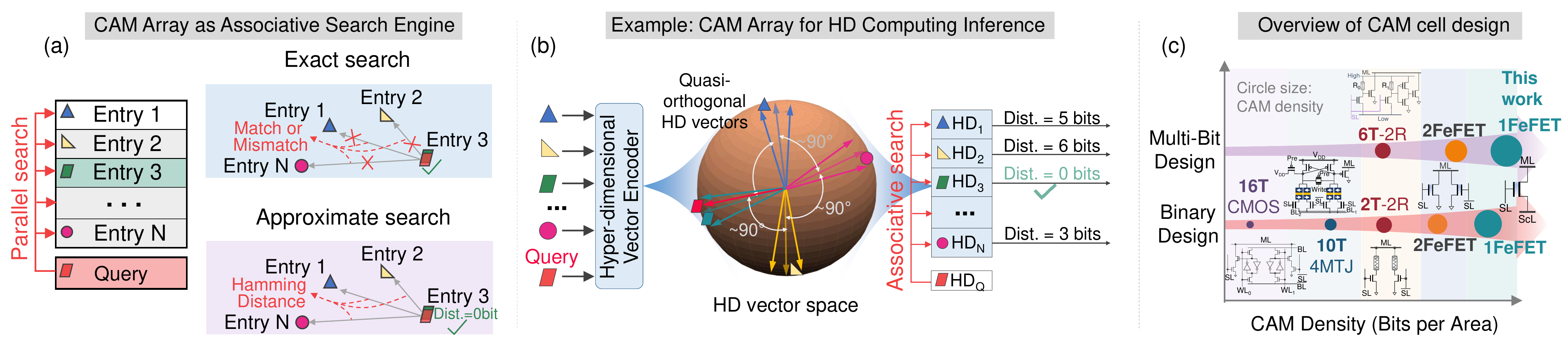}
	\vspace{-2ex}
	\caption{Overview of CAM based associative search engine for data-intensive pattern searching applications. (a) CAM array, due to its massive parallelism and in-memory computing capability, can search the input query either in the exact matching mode or approximate matching mode. (b) The approximate matching mode, which can calculate the distance between query and stored entries, can find wide applications, for example as associative memory in hyperdimensional computing. (c) Overview of existing BCAM and MCAM designs in terms of CAM bit density. The proposed 1FeFET based CAM design achieves the highest density by leveraging only one active memory device.}
	\label{fig:fig1_overview}
\end{figure*}

Besides matrix multiplications, the search operations are also prevalently seen and at the core of many applications,
and accelerating the searches over a class of data vectors can directly benefit various computational models and improve the system performance.
As a special form of IMC solutions, 
content addressable memories (CAMs) can accelerate parallel search operations throughout an entire memory array, thus demonstrating a promising potential utility in modern computing platforms \cite{hu2021in, ni2019ferroelectric, karam2015emerging}. CAM, depending on the stored value (i.e., binary, ternary, or multi-bit), can be classified as binary CAM (BCAM), ternary CAM (TCAM) (i.e., a third "don't care" or wildcard state), or multi-bit CAM (MCAM) \cite{hu2021in,yin2020fecam}. When given an input query, a CAM simultaneously compares each of its stored memory entry with the input, and returns the stored entries that match with the input, as shown in Fig. \ref{fig:fig1_overview}(a). The search can be performed in either exact mode or approximate mode. In the former scenario, only the entirely matched entries will be identified, whereas in the latter case, the distances (i.e., Hamming distance for BCAM/TCAM \cite{ni2019ferroelectric} while a novel distance metric for MCAM \cite{kazemi2021fefet}) between the stored entries and the input query are calculated, as shown in Fig. \ref{fig:fig1_overview}(a), thus serving as a distance kernel for various applications \cite{hu2021in, ni2019ferroelectric, kazemi2021fefet}.


One promising application that can significantly benefit from CAM is hyperdimensional computing (HDC), which can perform cognition tasks, such as image classification and speech recognition \cite{wu2018hyperdimensional, imani2019searchd, ge2020classification}. As a brain-inspired computing model, in HDC, class vectors are represented as almost orthogonal hypervectors in a high dimensional space (e.g., thousands of dimensions and each dimension is independent and identically distributed), as shown in Fig. \ref{fig:fig1_overview}(b) \cite{ge2020classification}. The HDC inference for classification is performed by identifying classes that are closest to the input query. 
CAM can greatly accelerate the HDC inference by storing class hypervectors (HD\textsubscript{N} in Fig. \ref{fig:fig1_overview}(b)) and calculating the distances between stored class hypervectors and input search query vector (HD\textsubscript{Q} in Fig. \ref{fig:fig1_overview}(b)) in memory and through a massively parallel fashion. 
HDC also finds wide applications beyond cognition tasks, taking genome sequencing as an example in this work. Genome sequencing is a typical pattern matching problem that is widely applied in bioinformatics applications. 
In this task, a genome sequence is searched through the overall genome library for entries that contain the query sequence. 
Despite of the importance, the efficient acceleration of the pattern matching for the genome sequencing is still an open question. 
HDC has been proposed as an effective solution as it can transform the inherent sequential processes of pattern matching to highly parallelizable computation tasks and translate the complex distance metric between the patterns to hamming distance \cite{kim2020geniehd}. 
These properties make the CAM an ideal platform to accelerate genome pattern matching via HDC.




To support the cognition  or genome sequencing tasks with growing complexity and data amount, having high density CAM arrays is indispensable. The traditional CMOS static random access memory (SRAM) based CAM design severely suffers from high area and power overheads due to a large number of transistors (i.e., 16 transistors) needed for a CAM cell and resulted large parasitics, thus limiting their practical use.
In order to address the aforementioned challenges faced by CMOS based BCAM/TCAM designs, two novel approaches have been proposed and studied, shedding light on further improving the system performance, area and energy efficiency over the CMOS based designs. 
One method is to leverage the emerging NVM devices to build compact and efficient BCAM/TCAM designs. As shown in Fig.\ref{fig:fig1_overview}(c), compact CAM designs can be achieved by using STT-MRAM (10T-4MTJ, MTJ: magnetic tunnel junction). To further improve the density, two-terminal resistive memories, such as ReRAM or phase change memory (PCM), can be used to construct 2T-2R BCAM/TCAM cell. The highest density CAM design so far is to leverage two FeFETs \cite{ni2019ferroelectric}.   

The second approach is to go beyond the conventional binary/ternary CAM designs and exploit the multi-level cell (MLC) property of NVMs for the design of MCAM. In a MCAM cell, multi-bit information is encoded, stored and searched, which provides an alternative route for information density boost. Such an approach is less explored compared with the BCAM designs, with only a few examples demonstrated so far. For example, leveraging the resistive memory devices, MCAM with 6T-2R, 
different from ReRAM based binary counterpart, 
has been proposed \cite{li2020analog}. An unique example is the 2 FeFET CAM cell, which has been shown to be a universal design that can simultaneously serve as a BCAM/TCAM cell and MCAM cell \cite{yin2020fecam}. That unique advantage of FeFET, along with its superior write energy-efficiency and density, make FeFET based CAM an excellent candidate for associative memory.    

All the aformentioned CAM designs, either binary or multi-bit, need at least two active memory devices. However, compact designs that use
only one memory element to achieve the ultimate CAM density, has yet been realized. Such a design, if existed, is also preferred to be universal, similar to the 2FeFET CAM cell \cite{yin2020fecam}, that can simultaneously function as a BCAM and MCAM. In this work, we propose an ultimate compact BCAM and MCAM design based on only 1 FeFET, leveraging its intrinsic 3-terminal transistor structure and nonvolatility. In the following, the FeFET device will first be explained and the 1FeFET universal CAM design will be proposed and validated. The proposed design will then be leveraged for the  genome sequencing application through HDC.     
\begin{figure*}[h]
	\centering
	\vspace{-1ex}
	\includegraphics [width=1.0\linewidth]{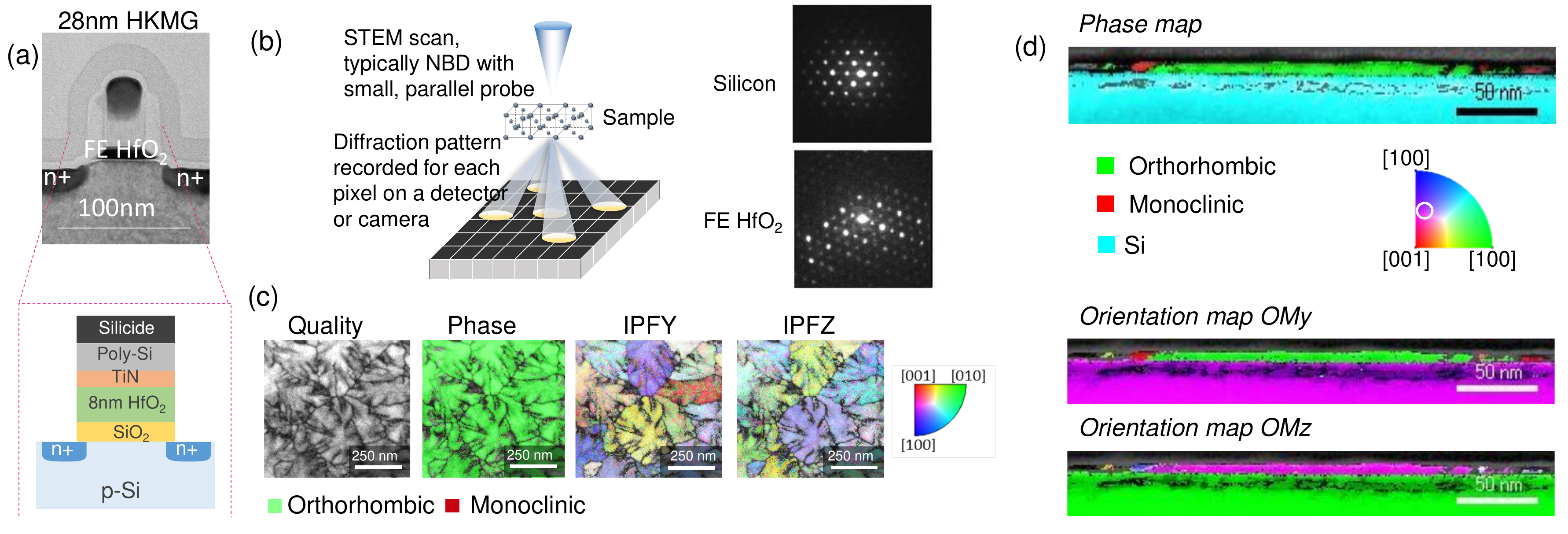}
	\vspace{-2ex}
	\caption{HfO$_2$ based FeFET. (a) Cross-sectional transmission electron microscopy (TEM) of the FeFET integrated into a 28nm high-k metal gate process and its schematic cross section. (b) TEM based electron diffraction characterization method used to identify with high spatial resolution the phase and orientation of grains by local indexation in integrated devices; reference diffraction patterns for silicon and ferroelectric HfO$_2$. (c,d) Phase map and inverse pole figure/orientation maps extracted after electron diffraction indexation of the  hafnium oxide grains in the high-k stack for an in-plane or cross-section view, respectively. An orientation close to [101] of the large polar orthorhombic grain in the cross-section can be extracted from OMz.}
	\label{fig:fig2_fefet_material_device}
\end{figure*}

\section{HfO$_2$ based FeFET}
\label{sec:device}

The discovery of ferroelectricity in doped HfO\textsubscript{2} \cite{boscke2011ferroelectricity} has triggered significant interests in its integration in FeFET for high-density and energy-efficient embedded NVM. By applying a positive/negative pulse on the gate, the ferroelectric polarization is set to point toward the channel/gate electrode, setting the FeFET threshold voltage, \textit{V}\textsubscript{TH}, to be either low/high value, respectively. The stored memory state can then be sensed through the channel current by applying a read gate bias between the low-\textit{V}\textsubscript{TH} (LVT) and high-\textit{V}\textsubscript{TH} (HVT) states. With its electric field driven write mechanism, FeFET exhibits superior write energy-efficiency, thus appealing for IMC applications. In this work, FeFET devices were fabricated using a 28 nm node gate-first high-k metal gate CMOS process on 300 mm silicon wafers. Detailed process information can be found in \cite{trentzsch201628nm}. As shown in Fig.\ref{fig:fig2_fefet_material_device}(a), the fabricated FeFET features a poly-crystalline Si/TiN (2 nm)/doped HfO\textsubscript{2} (8 nm)/SiO\textsubscript{2} (1 nm)/p-Si gate stack. The ferroelectric gate stack process module starts with growth of a thin SiO\textsubscript{2} based interfacial layer, followed by the deposition of an 8 nm thick doped HfO\textsubscript{2}. 
A TiN metal gate electrode was deposited using physical vapor deposition (PVD), on top of which the poly-Si gate electrode is deposited. The source and drain n$^+$ regions were obtained by phosphorous ion implantation, which were then activated by a rapid thermal annealing (RTA) at approximately 1000 $^\circ$C. This step also results in the formation of the ferroelectric orthorhombic phase within the doped HfO\textsubscript{2}.

As the microstructure of the ferroelectric layer plays a vital role for its application inside the FeFET memory, we analyzed the microstructure on a planar film before gate structuring using STEM with a dedicated detector with high dynamic range at each pixel, as shown in Fig.\ref{fig:fig2_fefet_material_device}(b). Similar to transmission Kikuchi diffraction (TKD)\cite{TKD_Lederer1} which has been utilized to investigate the granular structure of polycrystalline HfO$_2$ films, the analysis is carried out in transmission, with the detector located inside the beam. The main differences here are the accelerator voltage and an adjustable convergence angle, which allows for tuning the measured signal between classical electron diffraction and Kikuchi diffraction patterns \cite{TKD, TKD_Lederer2}. The former is used in this analysis and can be similar to TKD matched with extracted diffraction patterns for specific phases and crystal orientation for indexation. For reference, diffraction patterns for silicon and ferroelectric HfO$_2$ are shown in Fig.\ref{fig:fig2_fefet_material_device}(b). 
Based on the set electron diffraction images the phase and grain orientation can be indexed in the scan, which is shown with the phase and orientation maps in Fig. 2(c) and 2(d) for in-plane and cross-section, respectively. We can observe a highly polar orthorhombic phase fraction in the film. Still, a small monoclinic phase fraction can be identified, an area ratio of less than 5\% has been extracted. The orientation of the grains is extracted and shows a homogeneous orientation within the grains. A preferred tilted out-of-plane orientation along the <110> axes can be identified.
\begin{figure*}[h]
	\centering
	\vspace{-1ex}
	\includegraphics [width=1.0\linewidth]{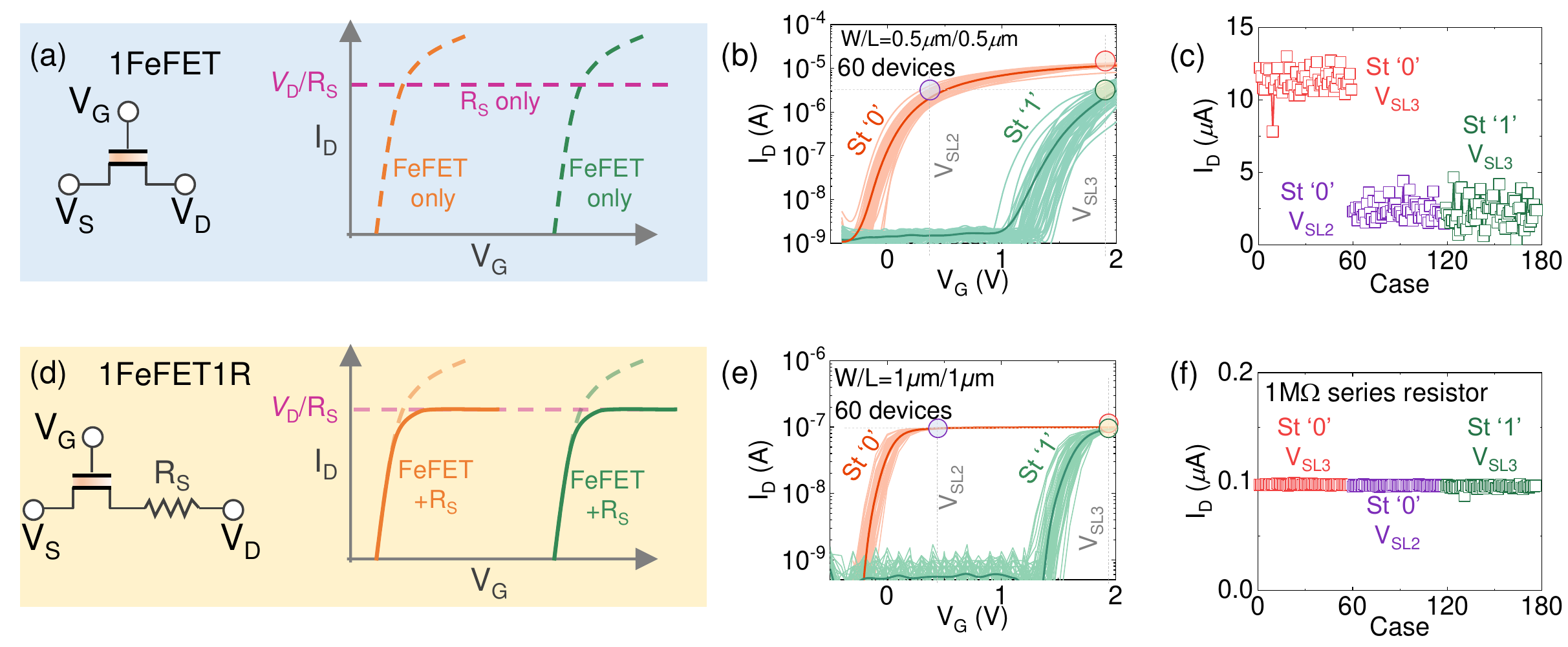}
	\vspace{-2ex}
	\caption{Suppression of \textit{I}\textsubscript{ON} variation and \textit{V}\textsubscript{G} dependency through a series current limiter. (a) A normal FeFET exhibits the conventional transistor characteristics. A constant current flows from source to drain if the transistor is replaced by a resistor \textit{R}\textsubscript{S}. (b) FeFET \textit{I}\textsubscript{D}-\textit{V}\textsubscript{G} for 60 different devices. From the characteristics, two gate bias, \textit{V}\textsubscript{SL2} to read state '0' and \textit{V}\textsubscript{SL3} to read state '1', can be chosen to yield the same average \textit{I}\textsubscript{ON}. (c) The corresponding read \textit{I}\textsubscript{ON} when applying \textit{V}\textsubscript{SL2} and \textit{V}\textsubscript{SL3} to both state '0' and '1'. Significant \textit{V}\textsubscript{G} dependence and variation of ON current is exhibited. With (d) a series current limiter (e.g., a resistor), (e) FeFET \textit{I}\textsubscript{D}-\textit{V}\textsubscript{G} curves is clamped by the series current limiter. (f) ON current is made independent of FeFET, but solely dependent on the applied drain bias and resistor. \textit{V}\textsubscript{D}=0.1 V for all the measurements. }
	\label{fig:fig3_1fefet1r}
\end{figure*}

\section{Proposal of 1FeFET CAM}
\label{sec:design}

\subsection{Target FeFET Characteristics for IMC}
\label{sec:targetdevice}

IMC, such as the crossbar array for matrix-vector multiplications or the CAM for  bit-wise XNOR operations, typically operates on the current domain, as shown in Fig.\ref{fig:figs1_device_for_imc} (a) and (b). 
In the implementation, the current from an array of memory devices are summing together, representing the output results of the intended operation. In such current mode of operation, it is critical that the ON current variability of the FeFET devices are well controlled such that the error rate induced by the current variation can be suppressed. For a conventional FeFET, the ON current variability of the device with a stored \textit{V}\textsubscript{TH} state is directly correlated with the \textit{V}\textsubscript{TH} variation, as shown in Fig.\ref{fig:figs1_device_for_imc}(c). Fig.\ref{fig:fig3_1fefet1r}(a) and (b) show the \textit{I}\textsubscript{D}-\textit{V}\textsubscript{G} characteristics of LVT and HVT states of 60 FeFETs. Non-negligible device-to-device variation is observed, which would cause significant variation in ON current, as shown in Fig.\ref{fig:fig3_1fefet1r}(c). 
Moreover, when the device is read with different gate biases (i.e., \textit{V}\textsubscript{SL2} and \textit{V}\textsubscript{SL3} in Fig.\ref{fig:fig3_1fefet1r}(b), which are chosen to turn on the device LVT and HVT states, respectively, and yield the same ON current), it can be observed that the ON current of the device with LVT is not constant, but highly dependent on the gate bias \textit{V}\textsubscript{G}.
This places a stringent requirement on controlling \textit{V}\textsubscript{TH} variation in FeFET, which is rapidly progressing with the optimization of integration process yet still faces challenges especially with FeFET scaling \cite{dunkel2017fefet, beyer2020fefet}. 
Two approaches are studied to address the variation challenges. The first straightforward method is continual process optimization, such as the poly-crystalline phase control, grain orientation control, etc. 

In this work, we  adopt an alternative method proposed in \cite{soliman2020ultra}, which is to limit the ON current variability by a limiter. In this way, the ON current of FeFET is effectively independent on both the applied \textit{V}\textsubscript{G} and the stored \textit{V}\textsubscript{TH} state, therefore, the \textit{V}\textsubscript{TH} variation of FeFET will not be translated into the ON current variation, as shown in Fig.\ref{fig:figs1_device_for_imc}(d). With significantly suppressed ON current variability, the accuracy and robustness of IMC solutions are greatly improved \cite{soliman2020ultra}. 
A straightforward implementation of the current limiter is a series resistor, as shown in Fig.\ref{fig:fig3_1fefet1r}(d). 
The ON current is determined by the component with a larger resistance value. The series resistor need to be greater than the effective turning-on channel resistance of FeFET  such that the ON current is simply dominated by \textit{V}\textsubscript{D}/\textit{R}\textsubscript{S} once FeFET turns ON. We performed experimentally device-to-device measurements on the 1FeFET1R structure \textit{I}\textsubscript{D}-\textit{V}\textsubscript{G} characteristics for 60 different devices, as shown in Fig.\ref{fig:fig3_1fefet1r}(e). The \textit{I}\textsubscript{D}-\textit{V}\textsubscript{G} characteristics show \textit{V}\textsubscript{G} independent ON state current with significantly suppressed variability,
which can be highly beneficial for IMC implementation. 
Extracted ON currents for the two read gate voltages, similar to Fig.\ref{fig:fig3_1fefet1r}(c), again confirms the effectiveness of the series current limiter in suppressing the ON current variability, as shown in Fig.\ref{fig:fig3_1fefet1r}(f). 
The current limiter is robust against FeFET scaling, as shown in Fig.\ref{fig:figs2_target_reliazation} (a) and (b). 
With the gate length, \textit{L}\textsubscript{G}, scaled down to 80 nm, the current limiter can suppress the ON current variability and the \textit{V}\textsubscript{G} dependence of ON current.

Integration of the series resistor with FeFET is necessary to fully exploit the benefit of current limiter. \cite{saito2021analog} proposed a TiN/SiO\textsubscript{2} tunneling junction based resistor integrated in the back-end-of-line (BEOL) and connected with FeFET drain. This approach, though effective, is inconvenient to implement. Another design is to adopt a split-gate structure, similar to the split-gate embedded FLASH memory \cite{richter2018cost}, where a conventional transistor is integrated in series with a FeFET. The conventional transistor can be used to generate gate-tunable series resistance, which adds additional flexibility in its application. But this design significantly complicates the integration processes. It would be desirable to design FeFET with a built-in current limiter. 
Here we propose two designs, one is to adopt Schottky source/drain contact (Fig.\ref{fig:figs2_target_reliazation}(c)) and the other one is to use underlapped channel region (Fig.\ref{fig:figs2_target_reliazation}(f)). As shown in Fig.\ref{fig:figs2_target_reliazation}(d) extracted from technology computer aided design (TCAD) simulation, once the transistor turns ON, the schottky barrier becomes the limiter for carrier transport. The simulated \textit{I}\textsubscript{D}-\textit{V}\textsubscript{G} characteristics at different barrier heights (Fig.\ref{fig:figs2_target_reliazation}(e)) clearly show the effectiveness of schottky barrier as current limiter. In the underlapped channel design, since it is not controlled by the gate, it can also function as a current limiter. Fig.\ref{fig:figs2_target_reliazation}(g) shows the simulated conduction band along the channel, which shows that the underlapped region limits carrier transport once the gate voltage is high. As a result, the \textit{I}\textsubscript{D}-\textit{V}\textsubscript{G} characteristics at different underlap lengths (Fig.\ref{fig:figs2_target_reliazation}(h)) show the effectiveness of underlapped region in limiting the ON current. These feasible designs are by no means optimal but  demonstrate that the FeFET based design for IMC implementation does not necessarily have to follow the conventional FeFET practices and the design space is large and remains to be explored.

\begin{figure*}[h]
	\centering
	\vspace{-1ex}
	\includegraphics [width=0.9\linewidth]{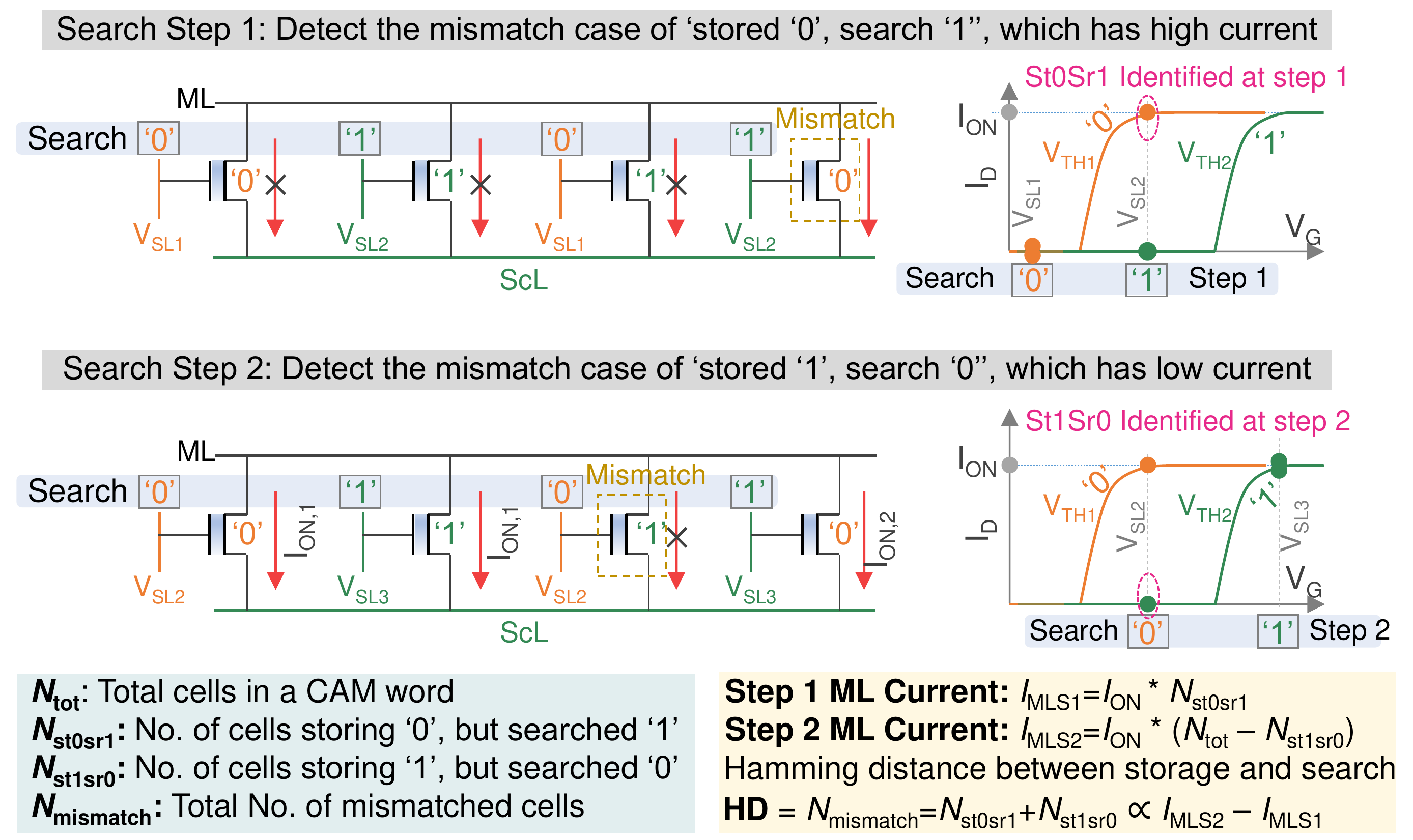}
	\caption{Two step search in the 1FeFET BCAM with \textit{V}\textsubscript{G} independent ON current. (a) The first step identifies the ‘St0Sr1’ mismatch, the only case with a high ML current in step 1. (b) The second step identifies the ‘St1Sr0’ mismatch, the only case with a low ML current in step 2. The difference between the two step ML current is proportional to the Hamming distance.}
	\label{fig:fig4_bcam_design}
\end{figure*}

\subsection{Demonstration of Binary CAM}
\label{sec:BCAM}

Thanks to its inherent transistor structure, a FeFET with \textit{V}\textsubscript{G} independent ON current can function as a BCAM or MCAM simultaneously. 
Such an ultra-compact CAM cell only consists of a single FeFET which can be fabricated from an 1FeFET1R structure as discussed in Sec. \ref{sec:targetdevice}. 
The proposed design leverages the property that a \textit{V}\textsubscript{TH} state stored in the FeFET (irrespective of actual \textit{V}\textsubscript{TH} value) can be uniquely identified with a two-step search scheme, where at the first step the search voltage is applied below \textit{V}\textsubscript{TH}, inducing negligible \textit{I}\textsubscript{OFF} leakage current, and in the second step the search voltage is applied above \textit{V}\textsubscript{TH}, conducting a high \textit{I}\textsubscript{ON} current.

With such a novel two-step search scheme, the FeFET \textit{V}\textsubscript{TH} state can be uniquely identified only when a low/high \textit{I}\textsubscript{D} flowing through the FeFET is sensed in the first/second search step, respectively.
By encoding the stored information into FeFET \textit{V}\textsubscript{TH} state and properly choosing the two-step search voltages (i.e., read \textit{V}\textsubscript{G}) to encode the query information, either a BCAM, leveraging the binary state of FeFET, with exact search and hamming distance calculation functions, or a MCAM, leveraging multi-level states of FeFET, with exact search function can be realized.
Below we illustrate the operation principles of BCAM design with 1FeFET structure per cell.

Fig.\ref{fig:fig4_bcam_design} shows the schematic of the 1FeFET CAM design along with our proposed two-step search scheme for the BCAM search function.
According to the device binary storage level characteristics, a state "0" can be written into the FeFET in a CAM cell by programming the device into LVT state, while a state "1" is written by programming the device into HVT state.
For the first step,  \textit{V}\textsubscript{SL1} below the \textit{V}\textsubscript{TH1} of LVT state is applied to the FeFET gate in the CAM cells to search bit "0", while \textit{V}\textsubscript{SL2} between the \textit{V}\textsubscript{TH1} of LVT and \textit{V}\textsubscript{TH2} of HVT is applied to search bit "1" as shown in Fig.\ref{fig:fig4_bcam_design}. 
For a matched search where the
FeFETs storing "0" are applied with \textit{V}\textsubscript{SL1} (denoted as St0Sr0) or the FeFETs storing "1" are applied with \textit{V}\textsubscript{SL2} (denoted as St1Sr1), since all search voltages (regardless of searching "0" or "1") are below the respective \textit{V}\textsubscript{TH}s, the current flowing through the matchline (ML) (i.e., \textit{I}\textsubscript{ML}) is negligible compared with the device on current \textit{I}\textsubscript{ON}.
The only mismatch condition that causes a high \textit{I}\textsubscript{ML} is when one or more FeFETs storing state "0" is searched with bit "1", i.e., applying \textit{V}\textsubscript{SL2} to the FeFETs in the LVT state (denoted as St0Sr1). Theoretically, the \textit{I}\textsubscript{ML} is linearly proportional to the number of St0Sr1 mismatch cases (denoted as \textit{N}\textsubscript{st0sr1}).
Therefore, the first search step identifies the St0Sr1 mismatch condition. For the exact search, any instance of St0Sr1 (\textit{N}\textsubscript{st0sr1} $\neq0$) means a mismatch, while for the approximate search where hamming distance is calculated, the ML current \textit{I}\textsubscript{ML} in this first step, \textit{I}\textsubscript{MLS1}, can used to calculate the number of mismatch bits, \textit{N}\textsubscript{st0sr1}.

For the second search step, the search voltages \textit{V}\textsubscript{SL2} and \textit{V}\textsubscript{SL3} above \textit{V}\textsubscript{TH2} of HVT state are applied to represent searching bit "0" and "1", respectively, as shown in Fig.\ref{fig:fig4_bcam_design}.
Unlike the first search step, there are three cases conducting the ON current \textit{I}\textsubscript{ON}: the FeFETs storing bit "0" are applied with \textit{V}\textsubscript{SL2} (i.e., St0Sr0), the FeFETs storing "1" are applied with \textit{V}\textsubscript{SL3} (i.e., St1Sr1), and the FeFETs storing "0" are applied with \textit{V}\textsubscript{SL3} (i.e., St0Sr1). 
The only mismatch case with a low \textit{I}\textsubscript{ML} is when applying \textit{V}\textsubscript{SL2} to the stored HVT state, i.e., FeFETs storing state "1" are searched with bit "0" (denoted as St1Sr0).
Therefore, sensing the step 2 ML current, \textit{I}\textsubscript{MLS2}, allows to detect how many mismatch cases of St1Sr0 (denoted as \textit{N}\textsubscript{st1sr0}) exist, as shown in Fig.\ref{fig:fig4_bcam_design}.
The total mismatch bits of the CAM word, i.e., the hamming distance between the stored word and the input query, can be calculated by adding aforementioned \textit{N}\textsubscript{st0sr1} and \textit{N}\textsubscript{st1sr0}, which is therefore linear to the current difference \textit{I}\textsubscript{MLS2} $-$ \textit{I}\textsubscript{MLS1}.

One can note that such BCAM design with our proposed two-step search scheme is impossible with the conventional FeFET devices, as shown in Fig.\ref{fig:figs3_failure_of_conventional_fefet}. The first step to detect the existence of St0Sr1 is robust since it is the only case conducting \textit{I}\textsubscript{ON}. 
The issue lies in the second search step to detect St1Sr0, where the cells upon St0Sr1 condition have a higher conducting current than the St0Sr0 and St1Sr1 conditions. 
Such \textit{I}\textsubscript{ON} difference may cause a strong overlap between the \textit{I}\textsubscript{MLS2}'s for different \textit{N}\textsubscript{st1sr0}, making it impossible to distinguish between different \textit{N}\textsubscript{st1sr0}, as illustrated in Fig.\ref{fig:figs3_failure_of_conventional_fefet}. This analysis therefore highlights the significance of current limiter in suppressing the \textit{V}\textsubscript{G} dependency and variation of ON current.



\begin{figure*}[h]
	\centering
	\vspace{-1ex}
	\includegraphics [width=0.9\linewidth]{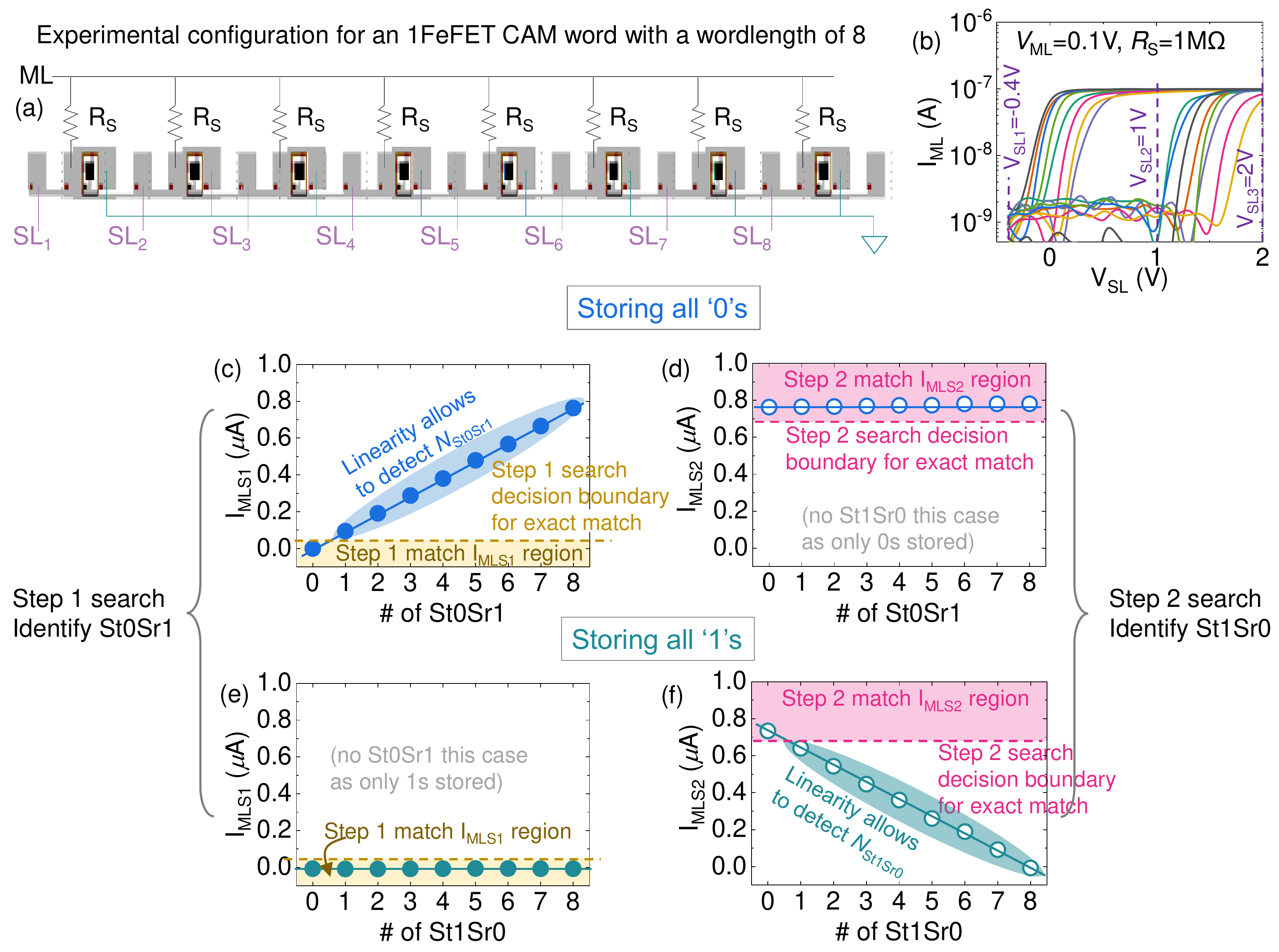}
	\caption{Experimental verification of 1FeFET CAM word with a wordlength of 8. (a) Experimental configuration for the measurement; (b) \textit{I}\textsubscript{D}-\textit{V}\textsubscript{G} of the 8 1FeFET1R devices. (c)/(d) and (e)/(f) the ML current at step 1/step 2 search when all cells stores ‘0’ and ‘1’, respectively. Successful array operation is demonstrated.}
	\label{fig:fig5_bcam_exp_verification}
\end{figure*}

The BCAM design has been validated through both experiments and simulations. Fig.\ref{fig:fig5_bcam_exp_verification} shows the experimental verification of the proposed BCAM array with a wordlength of 8.
Fig.\ref{fig:fig5_bcam_exp_verification}(a) shows the experiment setup of the CAM array, where a series resistor is connected with each FeFET in series as a current limiter. All the CAM cells are connected in parallel and the current-voltage characteristic of each cell is shown in Fig.\ref{fig:fig5_bcam_exp_verification}(b). 
Again, it shows that a sufficiently large memory window exists and \textit{I}\textsubscript{ON} is independent of \textit{V}\textsubscript{G} and its variation is suppressed. 

Two test scenarios are considered, where in case I, all the cells  store  bit "0" and in case II, all the cells  store  bit "1". 
Fig.\ref{fig:fig5_bcam_exp_verification}(c) and (d) show the ML currents for case I, \textit{I}\textsubscript{MLS1} and \textit{I}\textsubscript{MLS2}, corresponding to search step 1 and step 2, respectively. 
For the step 1 search, the \textit{I}\textsubscript{MLS1} increases linearly with the number of St0Sr1 (i.e., \textit{N}\textsubscript{St0Sr1}), which demonstrates the consistency with the design principle illustrated in Fig.\ref{fig:fig4_bcam_design}. Locating the decision reference between \textit{N}\textsubscript{St0Sr1}=0 and \textit{N}\textsubscript{St0Sr1}=1 would enable the exact match detection, and sensing the \textit{I}\textsubscript{MLS1} directly allows to detect the \textit{N}\textsubscript{St0Sr1}.
For the step 2 search, since the St1Sr0 condition is absent in case I, all the 8 cells will conduct an \textit{I}\textsubscript{ON}, resulting in a full match \textit{I}\textsubscript{MLS2} measurement as shown in Fig.\ref{fig:fig5_bcam_exp_verification}(d).
The measured \textit{I}\textsubscript{MLS1} for step 1 search and \textit{I}\textsubscript{MLS2} for step 2 search for the case II are shown in Fig.\ref{fig:fig5_bcam_exp_verification}(e) and (f), respectively. 
Since no St0Sr1 conditions exists for case II, the \textit{I}\textsubscript{MLS1} is low and below the exact match decision boundary, as shown in Fig.\ref{fig:fig5_bcam_exp_verification}(e). 
The \textit{I}\textsubscript{MLS2} exhibits a linear relation with the number of St1Sr0 (i.e., \textit{N}\textsubscript{St1Sr0}), therefore allowing the detection of the other mismatch cases in addition to the St0Sr1 condition in Fig.\ref{fig:fig5_bcam_exp_verification}(c). These measurements experimentally validate the proposed BCAM design. 

To go beyond the CAM array with 8 cells per word (due to limits of experimental setup), we have also performed SPICE Monte Carlo simulations of a scaled CAM array with a wordlength of 64 using an experimentally calibrated FeFET compact model \cite{ni2018circuit} which incorporates experimentally measured device variations shown in Fig.\ref{fig:fig3_1fefet1r}(b). 
For comparison, we simulated an 1FeFET CAM array with 2 cells per word without current limiter, as shown in Fig.\ref{fig:figs4_verification_1fefet_bcam_sim}(a). 
As illustrated in Fig.\ref{fig:figs3_failure_of_conventional_fefet}, the step 1 search is robust while step 2 search fails due to the \textit{V}\textsubscript{G} dependent ON current and its large variation. However, with the current limiter,  Fig.\ref{fig:figs4_verification_1fefet_bcam_sim}(b) even demonstrates the successful operation of a CAM array with 64 cells per word, showing a great promise of the proposed design.    

\begin{figure*}[h]
	\centering
	\vspace{-1ex}
	\includegraphics [width=0.9\linewidth]{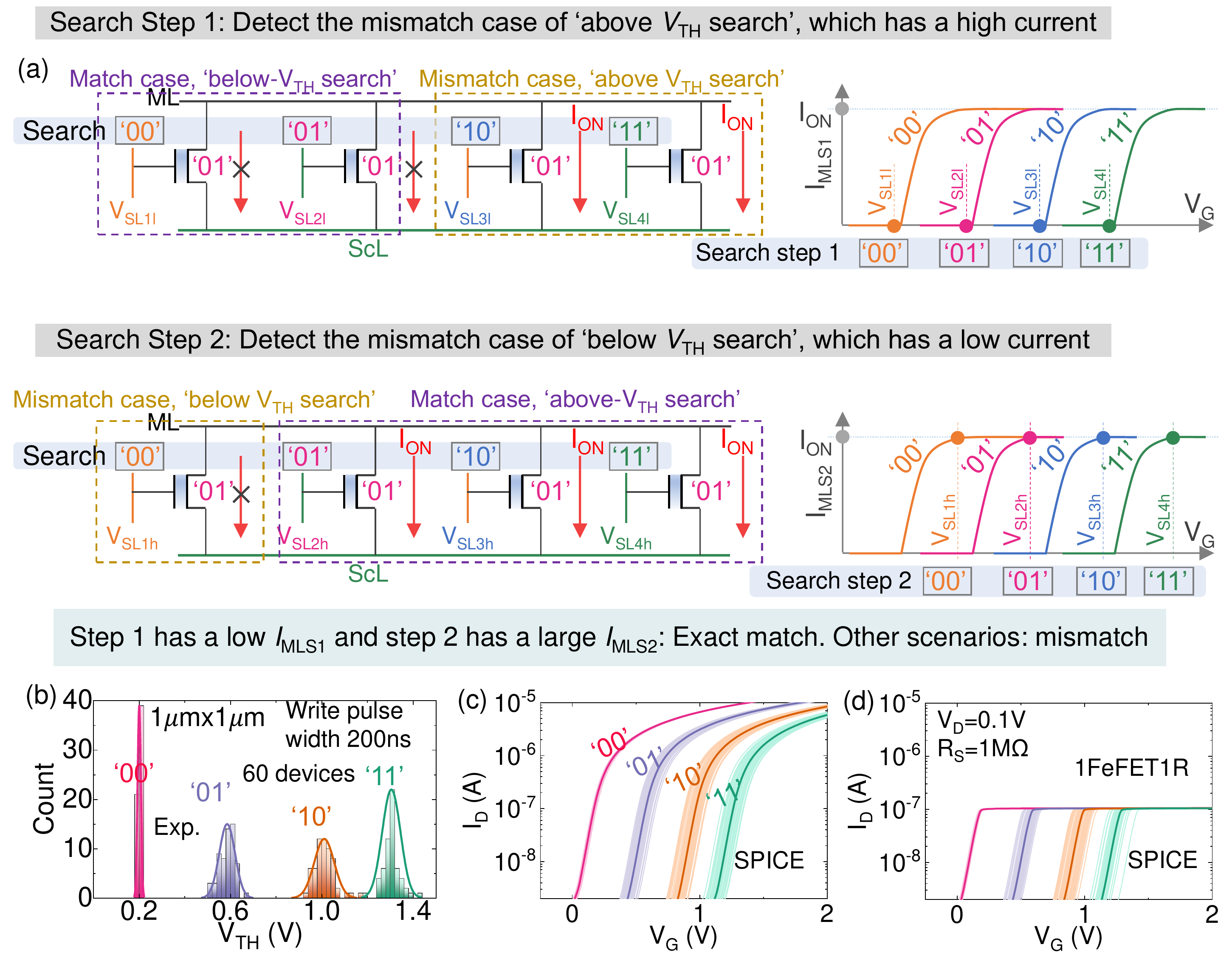}
	\vspace{-2ex}
	\caption{The 1FeFET CAM cell can also be used for MCAM. Two step exact search is executed. (a)/(b) Search step 1/step 2 performs ‘below-\textit{V}\textsubscript{TH}’/‘above-\textit{V}\textsubscript{TH}’ search, which detects ‘above-\textit{V}\textsubscript{TH}’/ ‘below-\textit{V}\textsubscript{TH}’ mismatch case, respectively. (c) 2 bits/cell measured in FeFETs. (d)/(e) SPICE simulation for 1FeFET/1FeFET1R with measured \textit{V}\textsubscript{TH} variation.}
	\label{fig:fig6_mcam_design_verification}
\end{figure*}

\subsection{Demonstration of Multi-Bit CAM}
\label{sec:MUltibitCAM}

As mentioned in Sec. \ref{sec:BCAM}, the proposed CAM cell design is universal as it can simultaneously serve as BCAM and MCAM cell. 
To determine whether the stored \textit{V}\textsubscript{TH}
state matches the encoded search information, a two-step search is proposed,  one below-\textit{V}\textsubscript{TH} search for detecting ‘above-\textit{V}\textsubscript{TH}’ mismatch cells and one above-\textit{V}\textsubscript{TH} search for detecting ‘below-\textit{V}\textsubscript{TH}’ mismatch cells. 
Only when all the cells match, the \textit{I}\textsubscript{MLS1} will be low in step 1 and the \textit{I}\textsubscript{MLS2} will be high in step 2. 
Any mismatch case of ‘above-\textit{V}\textsubscript{TH}’ search will be identified in step 1, and ‘below-\textit{V}\textsubscript{TH}’ mismatch search be detected in step 2. 
Following this search principle, we propose a MCAM with exact search function. 
Note that the proposed design methodology is scalable and theoretically applicable to FeFETs with any number of storable \textit{V}\textsubscript{TH} states. 

Fig.\ref{fig:fig6_mcam_design_verification} illustrates the operation principle using an example of 2 bits per cell.
Without loss of generality, considering all the cells storing bits "01", corresponding to the second lowest \textit{V}\textsubscript{TH} state, the search voltages representing the search bits "00", "01", "10", and "11" are applied to each cell, respectively. 
For step 1 search as shown in Fig.\ref{fig:fig6_mcam_design_verification}(a), a below-\textit{V}\textsubscript{TH} search, i.e., \textit{V}\textsubscript{SL3l} below the \textit{V}\textsubscript{TH} corresponding to state '10' and above the \textit{V}\textsubscript{TH} corresponding to state '01' is applied to search bit "10". 
In this way, any mismatch cell that is applied by a search voltage above the  \textit{V}\textsubscript{TH} corresponding to the stored state, i.e., 'above-\textit{V}\textsubscript{TH}' scenario (search bits "10" and "11" in this example), will conduct a high ON current \textit{I}\textsubscript{ON}. 
For step 2 search, a search voltage above-\textit{V}\textsubscript{TH} search, i.e., \textit{V}\textsubscript{SL3h} above the \textit{V}\textsubscript{TH} corresponding to state '10' and below the \textit{V}\textsubscript{TH} corresponding to state '11' is applied to search bit "10".
In this case, the only mismatched cells that are searched with a search voltage below the  \textit{V}\textsubscript{TH} corresponding to the stored state, i.e., 'below-\textit{V}\textsubscript{TH}' scenario (search bits "00" in this example), yield a low cell current. 
All the other cells conduct a high \textit{I}\textsubscript{ON}. 
Sensing the total ML current can then identify whether the 'below-\textit{V}\textsubscript{TH}' scenario, i.e., a mismatch cell exists. Combining the two step ML current results, the exact matched MCAM words can be identified as a low \textit{I}\textsubscript{MLS1} in step 1 and a high \textit{I}\textsubscript{MLS2} ($\approx$\textit{I}\textsubscript{ON}$\times$N, N is wordlength) in step 2. Any other ML current combinations indicate a mismatch.  

We also validated the design of MCAM, i.e.,  2 bits per cell using mixed experiments and SPICE simulations. 
Four \textit{V}\textsubscript{TH} states representing the stored 2 bits in FeFETs are demonstrated across 60 devices by leveraging the partial polarization switching of ferroelectric modulated with pulse amplitudes \cite{jerry2017ferroelectric}, as shown in Fig.\ref{fig:fig6_mcam_design_verification}(b).
The corresponding \textit{V}\textsubscript{TH} distributions are incorporated into the FeFET compact model \cite{ni2018circuit} and then Monte Carlo SPICE simulations are performed. 
Fig.\ref{fig:fig6_mcam_design_verification}(c) shows the FeFET \textit{I}\textsubscript{D}-\textit{V}\textsubscript{G} characteristics without the resistor current limiter. 
The ON current variability and \textit{V}\textsubscript{G} dependency for all the \textit{V}\textsubscript{TH} states are suppressed after incorporating the series current limiter. 
Such a cell enables the MCAM design. 
Fig.\ref{fig:figs5_mcam_array_sim} shows the simulation results on a MCAM word with 64 cells per word and 2 bits per cell. 
Without loss of generality, all the cells store bits "01" and the worst case search scenarios are considered, where only one cell mismatches with the search query. 
In Fig.\ref{fig:figs5_mcam_array_sim}(a), one cell is searched with bits "00", i.e., 'below-\textit{V}\textsubscript{TH}' scenario. 
This mismatch scenario is not seen in step 1 search, but shows up in the step 2 search. 
In Fig.\ref{fig:figs5_mcam_array_sim}(c) and (d), one cell is searched with bits "10" and "11", respectively, corresponding to 'above-\textit{V}\textsubscript{TH}' scenario. 
These mismatch scenarios are detected in step 1 search, which gives a high \textit{I}\textsubscript{ON}, and not discernible in step 2 search. 
Only when all the cells exactly match, as shown in Fig.\ref{fig:figs5_mcam_array_sim}(b), step 1 search will yield a low \textit{I}\textsubscript{MLS1} and step 2 search causes a high \textit{I}\textsubscript{MLS2} ($\approx$\textit{I}\textsubscript{ON}$\times$64), indicating a match scenario. Therefore, the results validate the proposed MCAM array function even with  current non-fully optimized FeFET.    
\section{Benchmarking and Application of 1FeFET CAM Array}
\label{sec:benchmarking_application}


\begin{figure*}[h]
	\centering
	\vspace{-1ex}
	\includegraphics [width=0.9\linewidth]{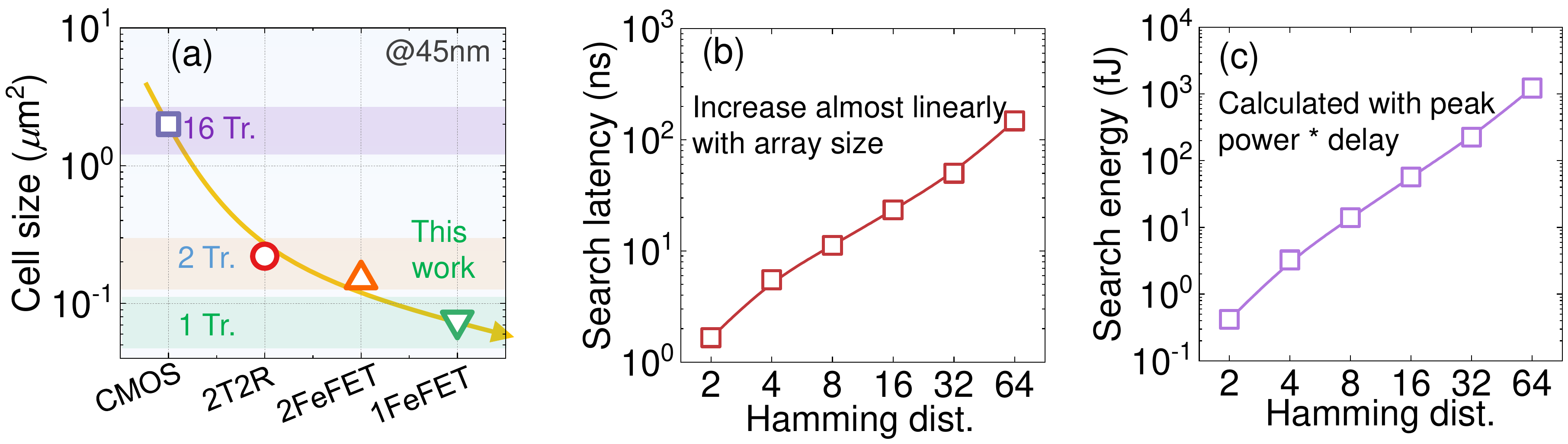}
	\vspace{-2ex}
	\caption{(a) The cell size comparisons of 1FeFET CAM with CMOS, 2T-2RRAM and 2FeFET CAM designs. The proposed cell is the most compact design. Search (b) latency and (c) energy of 1FeFET CAM array for search with different hamming distance thresholds.}
	\label{fig:fig7_array_benchmarking}
\end{figure*}

Fig.\ref{fig:fig7_array_benchmarking}(a) shows the CAM cell footprint scaling trend based on various emerging devices besides CMOS. 
Compared with existing CAM designs that consist of at least 2 devices per cell, our 1FeFET CAM design with the proposed two-step search scheme enables the most compact CAM cell area efficiency as well as the flexible BCAM and MCAM functionality. 
To sense the ML currents \textit{I}\textsubscript{ML} of our proposed CAM designs in the two-step search scheme, we adopt a thermometer-code analog digital converter (ADC) design that converts the ML currents to output voltages as the array sense amplifier, as shown in Fig.\ref{fig:figs6_adc}.
With the proposed design principles and proper reference currents, the output voltages of the sense amplifier design directly indicates the hamming distance between the input query and the stored words as illustrated in Fig.\ref{fig:fig4_bcam_design} when the 1FeFET CAM is configured as a BCAM, or indicates the match/mismatch condition as illustrated in Fig.\ref{fig:figs5_mcam_array_sim} when the 1FeFET CAM is configured as a MCAM. 
Since the sense amplifier is a serial design, the search delay and energy is proportional to the number of stages (i.e., word length), as shown in Fig.\ref{fig:fig7_array_benchmarking}. 
If the search delay is expected to be independent of the word length, other parallel designs for sensing can be adopted  yet at the cost of area and power overhead \cite{jiang2021analog}.

To evaluate the efficiency of the proposed 1FeFET CAM design beyond array level, we exploit the CAM array as an associative search engine for HDC running multiple genome sequencing tasks which are essential techniques in many bioinformatics applications.
In a genome sequencing task, a query DNA sequence represented by a string of nucleotide bases A, C, G, T, is searched in a reference DNA string which generally consists of 100 millions of DNA bases. Such pattern matching helps identify the existence of the query sequence in the reference sequence to discover potential diseases or accelerate DNA alignment techniques \cite{compeau2015bioinformatics, camacho2009blast+}.
Fig.\ref{fig:figs7_architecture}(a) shows the HDC architecture that efficiently parallelizes the genome sequencing tasks. In the architecture, the sequences from genome databases (E.coli, Human CHR14 and COVID-19) are encoded and stored in the associative memory (Fig.\ref{fig:figs7_architecture}(c)) for genome pattern matching.  
A new genome query is encoded and searched across multiple CAM banks in parallel, and the memory entries containing the sequences whose Hamming distances with the query are within a threshold are identified as illustrated in Fig.\ref{fig:figs7_architecture}(b). 
Fig.\ref{fig:figs7_architecture}(d) shows the benchmarking results of our proposed associative memory architecture for HDC genome sequencing tasks.
The results suggest that with the  hamming distance based approximate search associative memory implementation, our proposed search engine can achieve on average 89.9x (71.9x) faster and 66.5x (30.7x) higher energy efficiency as compared to state-of-the-art alignment tools NVBIO (GPU-BLAST) \cite{vouzis2011gpu} approach.

\section{Conclusion}
\label{sec:conclusion}

We proposed an ultra-compact and scalable 1FeFET CAM design with enhanced search function and improved CAM density for low power pattern matching application via HDC paradigm.
We fabricated the 1FeFET1R structure that integrates the series resistor into FeFET to limit the ON state current fluctuation, and experimentally demonstrated the 1FeFET CAM array with a novel two-step search operations.
The ultra-compact design has been demonstrated to perform hamming distance computation as a binary CAM for approximate search, and exact search function as a multi-bit CAM for improved CAM density. 
In addition to the high density, significant improvements in latency and energy efficiency have also been demonstrated when compared with the state-of-the-art hardware, highlighting the great promise of the proposed 1FeFET CAM design as an associative search memory engine.


\bibliography{ref}
\section{Conflict of Interest}
\label{sec:conflict}
The authors declare no conflicts of interest.

\newpage
\renewcommand{\thefigure}{S\arabic{figure}}
\renewcommand{\thetable}{S\arabic{table}}

\onecolumn
\centering
\textbf{\Large Supplementary Information}
\setcounter{figure}{0}
\setcounter{table}{0}
\setcounter{page}{1}
\begin{flushleft} 
\textbf{\large Target FeFET Characteristics for Current-Based In-Memory Computing}
\end{flushleft}

\begin{figure*}[h]
	\centering
	\vspace{-1ex}
	\includegraphics [width=1.0\linewidth]{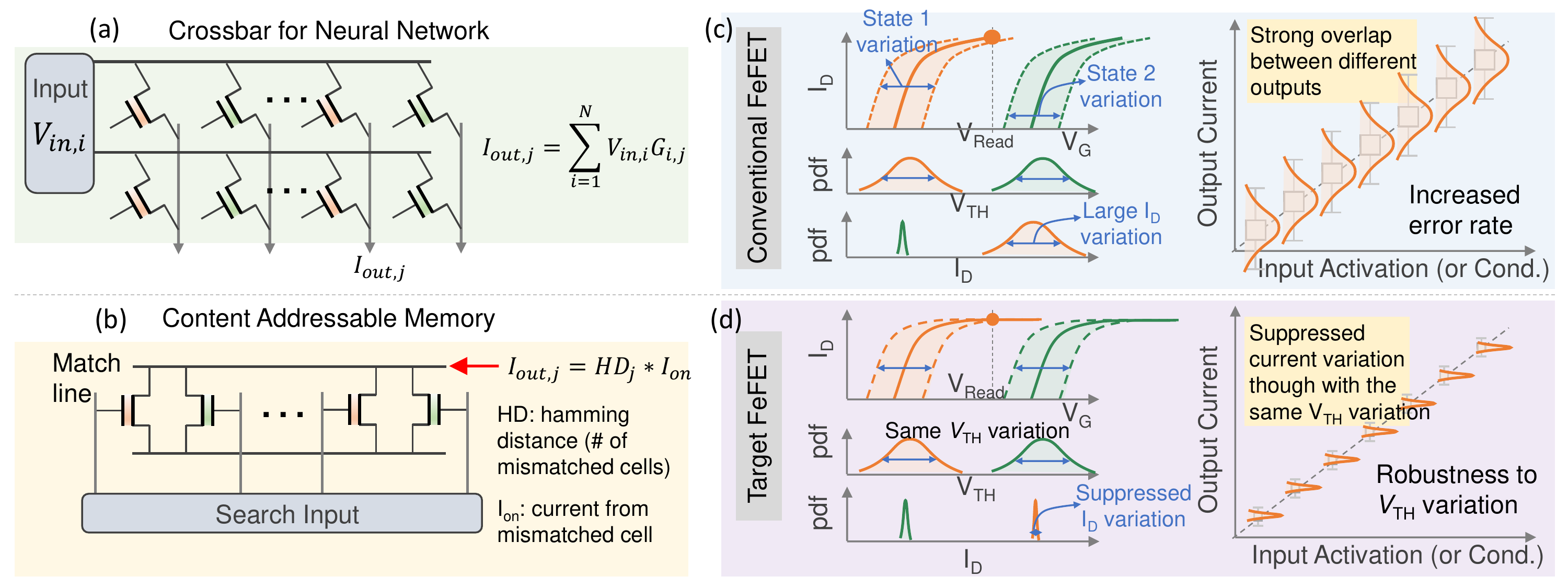}
	\vspace{-2ex}
	\caption{For current-based in-memory computing (IMC) elements, i.e., (a) crossbar for neural network and (b) content addressable memory for hamming distance computation, (c) the device's stored \textit{V}\textsubscript{TH}  state variability  and the ON current \textit{I}\textsubscript{D} variation are correlated. With large \textit{I}\textsubscript{D} variation due to the \textit{V}\textsubscript{TH} variation, strong overlaps between different output currents result in increased error rate being used in IMC circuits. (d) A device with \textit{V}\textsubscript{G}  and \textit{V}\textsubscript{TH} independent ON current is desirable to minimize the ON current variation. With ON current that is robust to \textit{V}\textsubscript{TH} variation, different output currents can be distinguished when being used in IMC circuits.}
	\label{fig:figs1_device_for_imc}
\end{figure*}

\justify
For current based in-memory computing using FeFETs, such as crossbar array for matrix-vector multiplication (multiply and accumulation, MAC) operation (Fig.\ref{fig:figs1_device_for_imc}(a)) and the content addressable memory for associative search (Fig.\ref{fig:figs1_device_for_imc}(b)), it is highly desirable to have a tight distribution of the ON current for FeFETs. For conventional FeFETs, the device current variation is closely related to the transistor \textit{V}\textsubscript{TH} variation. Especially for the ON current, the \textit{V}\textsubscript{TH} variation has an one-to-one correspondence to the \textit{I}\textsubscript{D} variation, as shown in Fig.\ref{fig:figs1_device_for_imc}(c). The ON current variability may induce output overlaps, thus sensing error, which could degrade the operation accuracy. 
This can be addressed through continual FeFET material and process optimization  to control both extrinsic sources and intrinsic polarization switching sources that cause \textit{V}\textsubscript{TH} variation. 
That said, new cell structure can be designed such that the cell ON current is independent of the \textit{V}\textsubscript{G} and \textit{V}\textsubscript{TH}, which is adopted in this work. 
Using a series current limiter on the FeFET drain, once the FeFET turns ON, the \textit{I}\textsubscript{D} is independent of the \textit{V}\textsubscript{G}  and \textit{V}\textsubscript{TH}, as shown in Fig.\ref{fig:figs1_device_for_imc}(d). This simple design, with its significantly suppressed \textit{I}\textsubscript{D} variation, can enable various applications, such as the 1FeFET CAM proposed in this work. 

\newpage
\begin{flushleft} 
\textbf{\large Realization of Target FeFET}
\end{flushleft}

\justify
The series current limiter can be applied for FeFETs with different sizes, as long as the equivalent ON resistance of FeFET is much less than the current limiter, so that the cell ON current is dominated by the current limiter and independent of the \textit{V}\textsubscript{G}. 

The series current limiter can be integrated into FeFET structure such that a true 1FeFET CAM cell can be realized. 
One intermediate design would be the split-gate structure typically used in embedded NOR flash transistor. However, in our design, the purpose of the series transistor is to limit the channel current. Also, the gate tunable resistance provides additional flexibility, which can be leveraged for different function demonstration, such as distance kernel for multi-bit CAM. 

\begin{figure*}[h]
	\centering
	\vspace{-1ex}
	\includegraphics [width=1.0\linewidth]{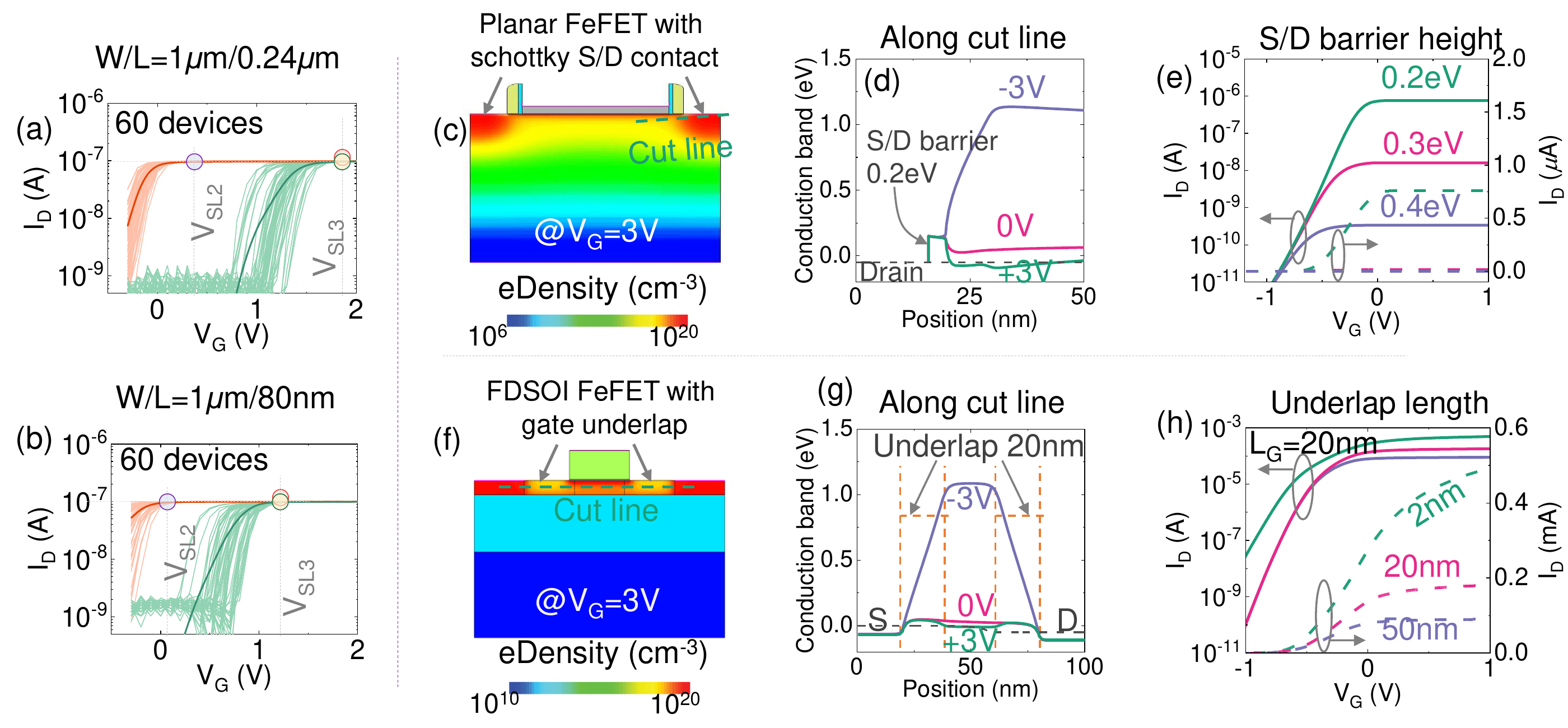}
	\vspace{-2ex}
	\caption{FeFET \textit{I}\textsubscript{D}-\textit{V}\textsubscript{G} curves with a series current limiter on the drain for (a) W/L=1$\mu$ m/0.24$\mu$ m and (b) W/L=1$\mu$ m/80 nm. Possible designs that can integrate the series current limiter into the FeFET structure are simulated in TCAD. (c) Planar FeFET with schottky source/drain contact. (d) Conduction band diagram along the cut line in (c) at different gate voltages. Once the FeFET turns on, the source/drain schottky barrier becomes the limit for carrier transport. (e) The \textit{I}\textsubscript{D}-\textit{V}\textsubscript{G} curves for different schottky barrier heights show the \textit{V}\textsubscript{G} independent ON current. (f) FDSOI FeFET with gate underlap. (g) Conduction band diagram along the channel for different gate biases. Once the gate voltage is high enough, the underlap regions become the limit for carrier transport. (h) The \textit{I}\textsubscript{D}-\textit{V}\textsubscript{G} curves for different underlap length. The longer the underlap length, the more independent ON current on the \textit{V}\textsubscript{G}.}
	\label{fig:figs2_target_reliazation}
\end{figure*}

\begin{figure*}[h]
	\centering
	\vspace{-1ex}
	\includegraphics [width=0.7\linewidth]{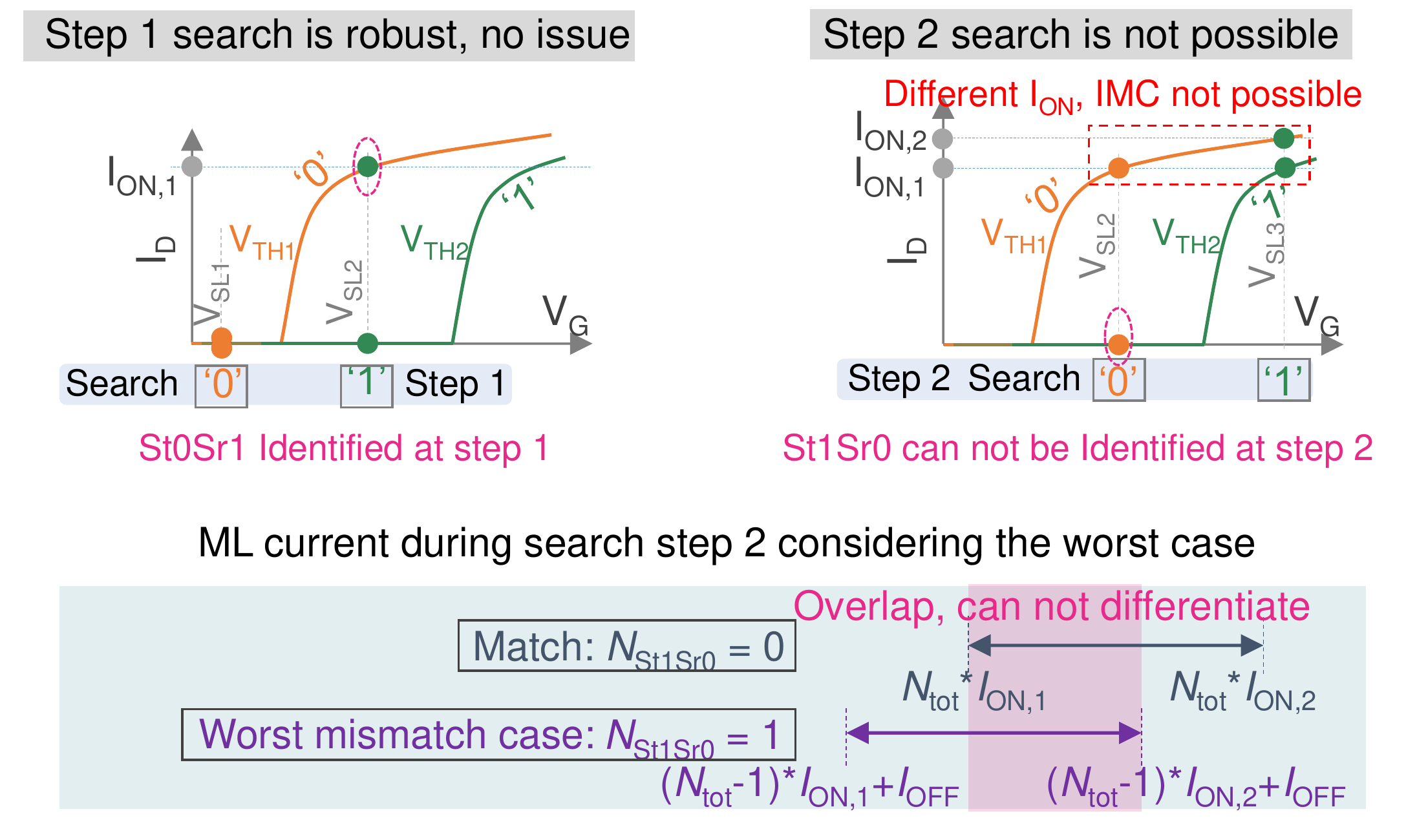}
	\vspace{-2ex}
	\caption{A FeFET with \textit{V}\textsubscript{G}-dependent \textit{I}\textsubscript{ON} cannot identify the 'St1Sr0' mismatch in step 2 search due to the overlap of the ML current between different \textit{N}\textsubscript{St1Sr0} cases.}
	\label{fig:figs3_failure_of_conventional_fefet}
\end{figure*}

\begin{figure*}[h]
	\centering
	\vspace{-1ex}
	\includegraphics [width=1\linewidth]{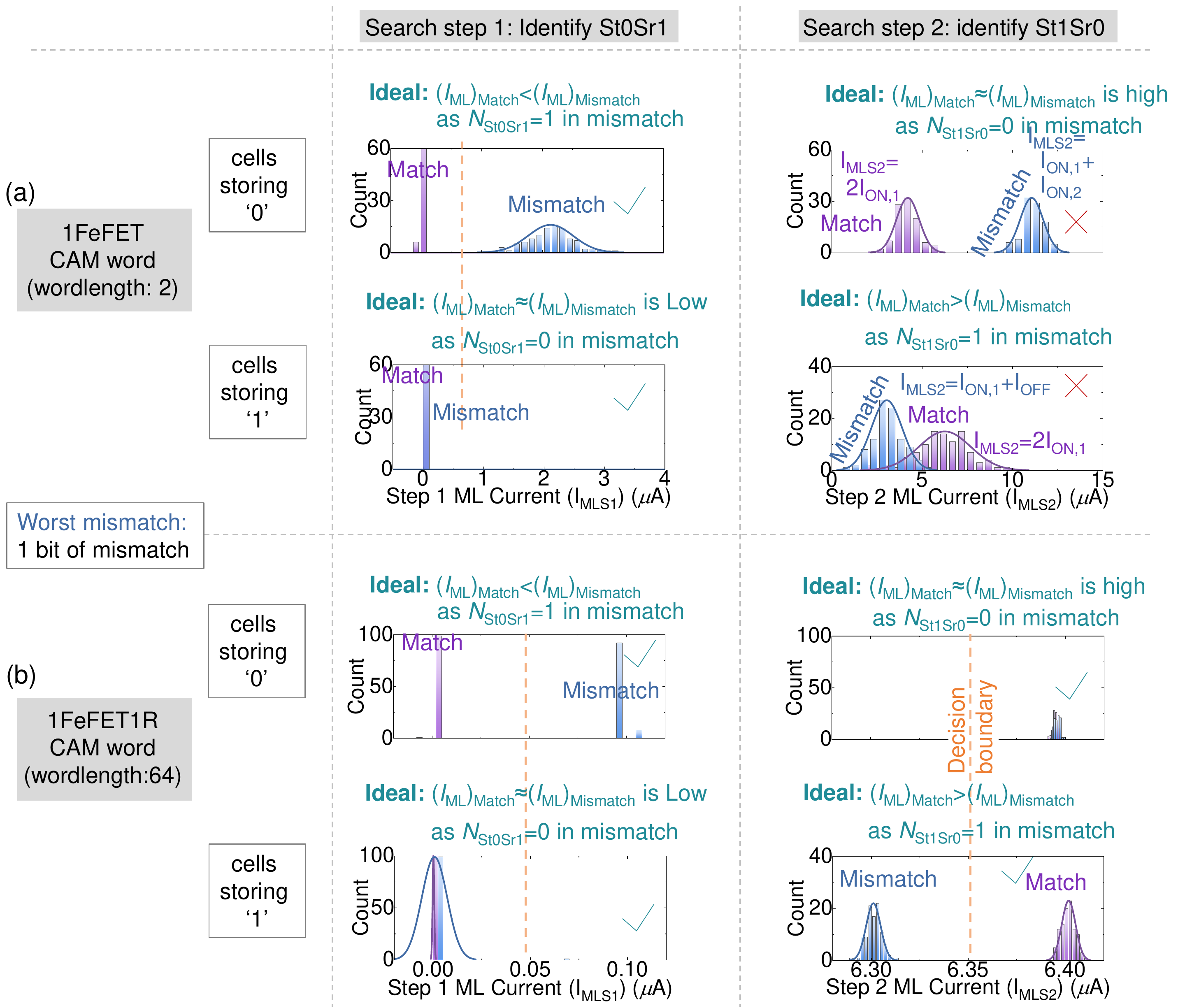}
	\vspace{-2ex}
	\caption{Verification of the 1FeFET BCAM word with SPICE simulation. Measured device-to-device variation is incorporated in the simulation. (a) The ML current distribution of a  BCAM word with a wordlength of 2 using conventional 1FeFET. Due to the strong \textit{I}\textsubscript{ON} variation and the \textit{V}\textsubscript{G} dependent \textit{I}\textsubscript{ON}, the step 2 search fails even for a wordlength of 2. (b) In contrast, for a BCAM word with a wordlength of 64 using 1FeFET with the resistor current limiter, successful operation can be demonstrated. }
	\label{fig:figs4_verification_1fefet_bcam_sim}
\end{figure*}

\begin{figure*}[h]
	\centering
	\vspace{-1ex}
	\includegraphics [width=0.95\linewidth]{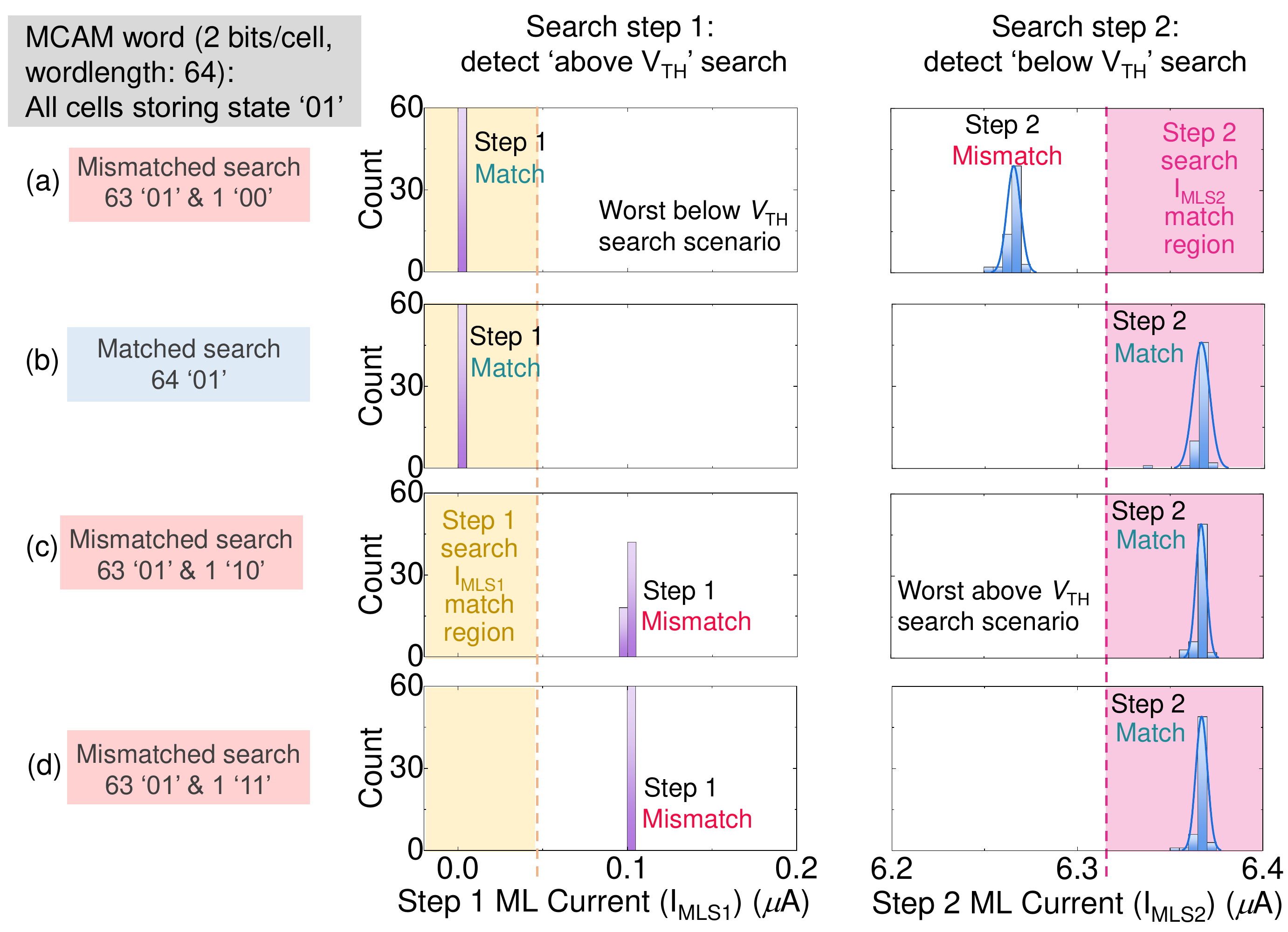}
	\vspace{-2ex}
	\caption{Verification of the 1FeFET MCAM on a MCAM word with a wordlength of 64. The case where all the cells storing state '01' is considered as an example. Worst cases where only 1 cell mismatch are considered. Different search scenarios where (a) 63 matched cells with input '01', 1 mismatched cell with input '00'; (b) fully matched cells; (c) 63 matched cells with input '01', 1 mismatched cell with input '10'; and (d) 63 matched cells with input '01', 1 mismatched cell with input '11' are considered. Validated successful operations are demonstrated.}
	\label{fig:figs5_mcam_array_sim}
\end{figure*}

Other designs are also considered here. One is to incorporate the Schottky barrier in the source/drain, as shown in Fig.\ref{fig:figs2_target_reliazation}(c). From TCAD simulations, the conduction band diagrams at different \textit{V}\textsubscript{G}, shown in Fig.\ref{fig:figs2_target_reliazation}(d),  show that once the channel turns ON, the current transport barrier is dominated by the Schottky barrier, which is \textit{V}\textsubscript{G} independent. Therefore the \textit{I}\textsubscript{D}-\textit{V}\textsubscript{G} curves shown in Fig.\ref{fig:figs2_target_reliazation}(e) show that the ON current is independent of the \textit{V}\textsubscript{G}.

Another proposed design  is the underlap FDSOI FeFET, as shown in Fig.\ref{fig:figs2_target_reliazation}(f), where an ungated channel is inserted between source/drain and the gated channel. Because this region is ungated, when the gate region turns on, the carrier transport will be limited by the ungated region, as shown in the conduction band diagram in Fig.\ref{fig:figs2_target_reliazation}(g). The \textit{I}\textsubscript{D}-\textit{V}\textsubscript{G} curves shown in Fig.\ref{fig:figs2_target_reliazation}(h) demonstrate that the longer the underlap length, the more effective in suppressing the dependence of ON current on \textit{V}\textsubscript{G}.  

\newpage
\begin{flushleft} 
\textbf{\large Failure of Conventional 1FeFET based CAM}
\end{flushleft}

\justify
For the proposed two-step search scheme, a conventional 1FeFET features an \textit{V}\textsubscript{G}-dependent \textit{I}\textsubscript{ON} characteristic. 
For the step 1 search, the number of  mismatch case 'St0Sr1' \textit{N}\textsubscript{St0Sr1} can be calculated by sensing the ML current which equals to \textit{N}\textsubscript{St0Sr1}$\times$\textit{I}\textsubscript{ON,1}.
For the step 2 search, the number of mismatch 'St1Sr0' \textit{N}\textsubscript{St1Sr0} cannot be calculated as different \textit{I}\textsubscript{ON}, i.e., \textit{I}\textsubscript{ON,1} and \textit{I}\textsubscript{ON,2} are sensed.
In this scenario, the ML current possibly includes \textit{I}\textsubscript{ON,1} corresponding to the match condition 'St0Sr0', 'St1Sr1' and \textit{I}\textsubscript{ON,2} corresponding to the mismatch condition 'St0Sr1'.
Therefore, 
upon a match case for the step 2 search,
the sensed ML current varies between \textit{N}\textsubscript{tot}*\textit{I}\textsubscript{ON,1} (only 'St0Sr0', 'St1Sr1' exist) and \textit{N}\textsubscript{tot}*\textit{I}\textsubscript{ON,2} (only 'St0Sr1' exists). In the worst mismatch case, where \textit{N}\textsubscript{St1Sr0}=1, the sensing current varies between (\textit{N}\textsubscript{tot}-1)*\textit{I}\textsubscript{ON,1}+\textit{I}\textsubscript{OFF} (only 'St0Sr0', 'St1Sr1' exist) and (\textit{N}\textsubscript{tot}-1)*\textit{I}\textsubscript{ON,2}+\textit{I}\textsubscript{OFF} (only 'St0Sr1' exists). 
The overlap between the two ML current ranges disables the conventional FeFET from working as a CAM cell.

\newpage
\begin{flushleft} 
\textbf{\large Verification of the Proposed 1FeFET Binary CAM}
\end{flushleft}

\justify
The 1FeFET BCAM operation is further verified using SPICE simulations. The experimental FeFET device-to-device variation shown in Fig.\ref{fig:fig3_1fefet1r}(b) is adopted for the Monte Carlo simulations. We consider two scenarios where all the cells storing bit '0' and all the cells storing bit '1', respectively.
The worst mismatch is considered where only 1 bit mismatch needs to be detected. Due to the strong variation in \textit{I}\textsubscript{ON} and \textit{V}\textsubscript{G} dependent \textit{I}\textsubscript{ON} in the conventional FeFET, even a small CAM word (i.e., wordlength of 2) operation fails in step 2 search. 
With the series current limiter, i.e., resistor, on the drain of FeFET, the variation in \textit{I}\textsubscript{ON} is significantly suppressed. 
As a result, a clear boundary can be defined between different degrees of mismatch, allowing to detect the hamming distance reliably.   

\newpage
\begin{flushleft} 
\textbf{\large Verification of the Proposed 1FeFET Multi-Bit CAM}
\end{flushleft}

\justify
The exact match functionality of 1FeFET MCAM is also verified on a word with a wordlength of 64 via SPICE simulations . All the cells store state '01' and the worst mismatch cases where only 1 cell mismatches are considered. Monte Carlo simulations with realistic FeFET variation are performed. Results show that only for fully matched search, the step 1 current is low and the step 2 current is high. Other mismatched cases can not satisfy the match criteria in both search steps at the same time. 

\newpage
\begin{flushleft} 
\textbf{\large Sense Amplifier of the 1FeFET Array}
\end{flushleft}

\begin{figure*}[h]
	\centering
	\vspace{-1ex}
	\includegraphics [width=0.6\linewidth]{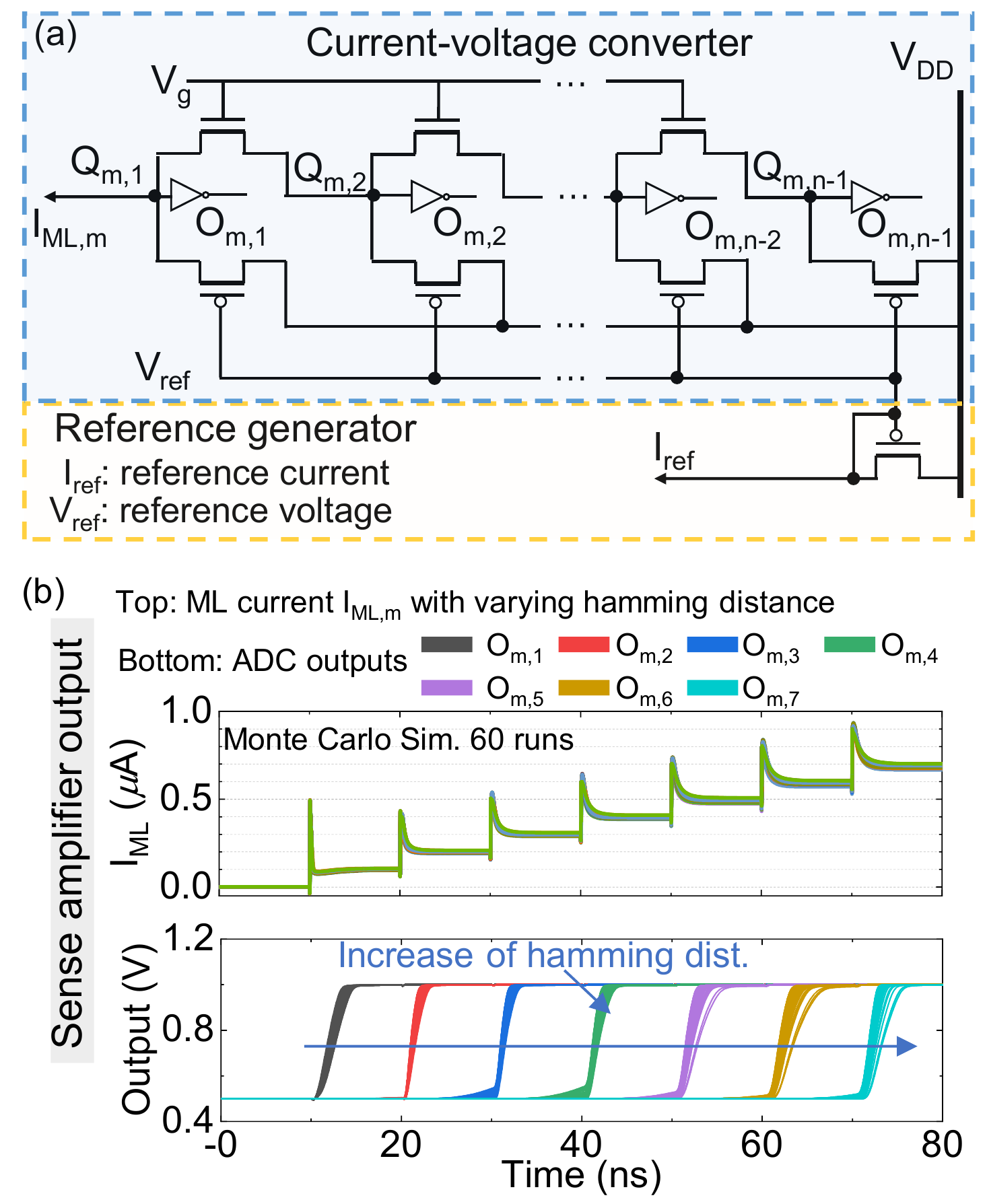}
	\vspace{-2ex}
	\caption{Sensing of the ML current. (a) The sense amplifier that can sense the ML current in a serial manner. (b) Simulated output waveforms for the sense amplifier.}
	\label{fig:figs6_adc}
\end{figure*}

\justify
The sense amplifier converts the ML current from the CAM array to the voltages, which indicate the hamming distance between the input query and the stored CAM word. 
The sense amplifier (Fig.\ref{fig:figs6_adc}(a)) consists of a reference  generator and a ladder-style current-voltage converter per word which connects with the ML of a CAM word. 
The reference generator generates the reference current \textit{I}\textsubscript{ref}, and transfers the reference voltage \textit{V}\textsubscript{ref} to the PMOS transistor gates of the stacked current mirrors, such that the PMOS transistors each can conduct \textit{I}\textsubscript{ref} when operating in saturation.
The serial connected conducting NMOS transistors of the current-voltage converter are biased at \textit{V}\textsubscript{G}.
During the search, the ML current \textit{I}\textsubscript{ML,m} is drawn by a CAM word, 
if \textit{I}\textsubscript{ML,m} is smaller than \textit{I}\textsubscript{ref}, the current mirrors maintains a high voltage at the PMOS drains, i.e., \textit{Q}\textsubscript{m,1},  \textit{Q}\textsubscript{m,2}, ..., \textit{Q}\textsubscript{m,n-1}, as no current is conducted in the NMOS transistor branch.
Once \textit{I}\textsubscript{ML,m} grows beyond \textit{I}\textsubscript{ref},  since the PMOS transistor in the leftmost current mirror can only provide \textit{I}\textsubscript{ref}, the voltage at \textit{Q}\textsubscript{m,1} quickly drops down and turns on its associated NMOS transistor to conduct the  extra current of \textit{I}\textsubscript{ML,m}, i.e., \textit{I}\textsubscript{ML,m}-\textit{I}\textsubscript{ref}, resulting in a high level output at \textit{O}\textsubscript{m,1}.
If \textit{I}\textsubscript{ML,m} keeps increasing and exceeds 2$\times$ \textit{I}\textsubscript{ref}, the voltage at \textit{Q}\textsubscript{m,2} drops down too, resulting in a high level output at \textit{O}\textsubscript{m,2}.
Similarly, such ladder-like current behavior repeats for the all subsequent stages as \textit{I}\textsubscript{ML,m} grows, and the simulation waveforms in Fig.\ref{fig:figs6_adc}(b) show that the outputs can detect the hamming distance of the CAM array as desired, depending on the required sense amplifier precision.

\newpage
\begin{flushleft} 
\textbf{\large 1FeFET CAM Array for Bioinformatics (e.g., DNA Genome Sequencing)}
\end{flushleft}

\justify
As an example of 1FeFET CAM application benchmarking, DNA genome sequencing through  hyperdimensional computing (HDC) is considered. 
HDC has been proposed as an effective solution for genome sequencing as it can transform the inherent sequential processes of pattern matching to highly parallelizable computation tasks and translate the complex distance metric to hamming distance.

\begin{figure*}[h]
	\centering
	\vspace{-1ex}
	\includegraphics [width=0.95\linewidth]{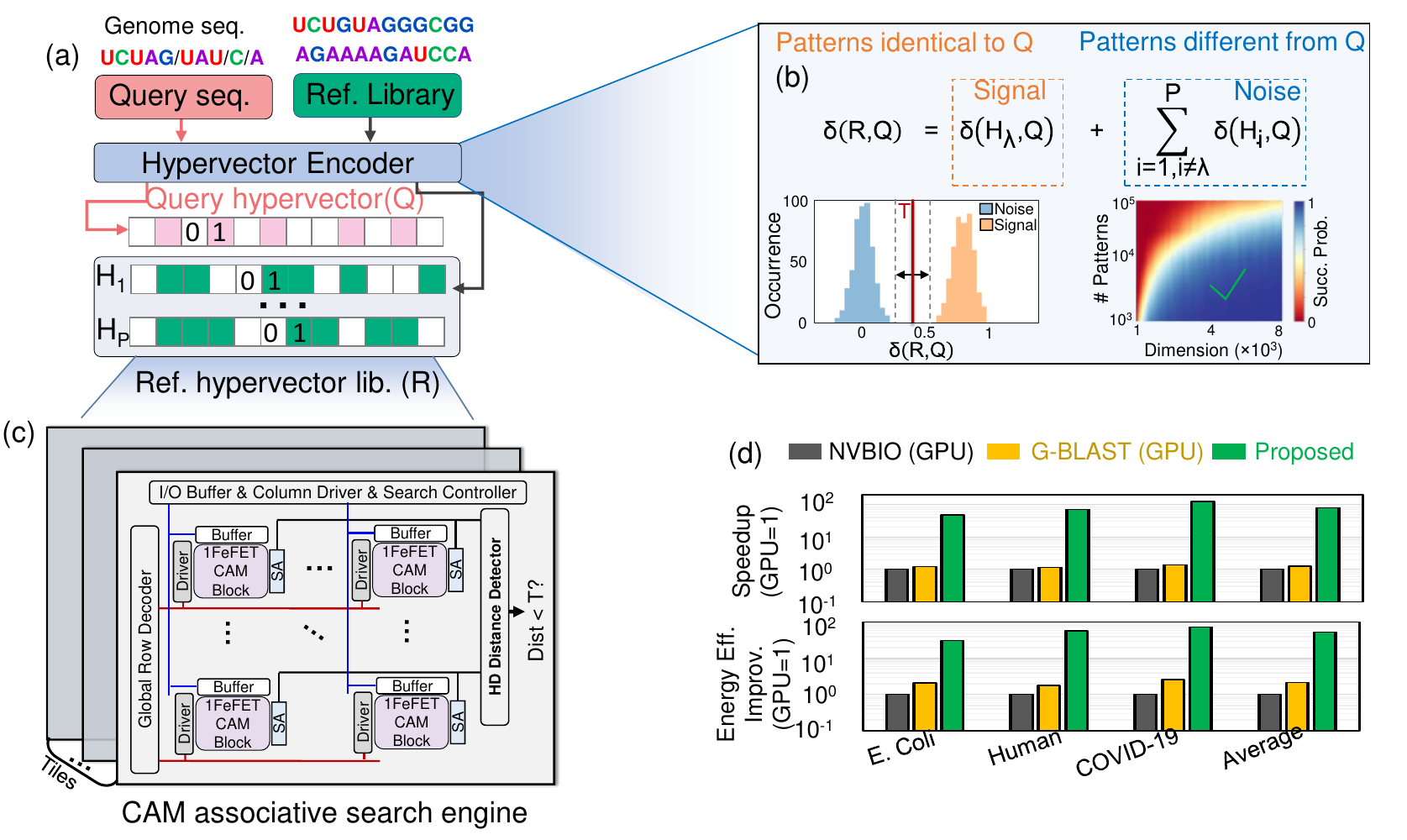}
	\vspace{-2ex}
	\caption{Benchmarking the 1FeFET CAM in the HDC system for DNA genome sequencing tasks. 
	(a) Overall flow of HDC for DNA genome sequencing. The genome library is encoded into random binary hypervectors, which are stored in the 1FeFET CAM. A query sequence goes through the same encoder to generate the query hypervector, which is then applied to the 1FeFET CAM to search for the entries in the reference library within a given distance threshold. (b) The hypervector encoder can map the genome sequence into almost orthogonal hypervectors such that the library entries that are identical to the query hypervector have much higher similarity than the library entries that are different. (c) 1FeFET CAM array architecture. (d) The 1FefET CAM structure can provides significant speedup and energy saving compared with GPU.}
	\label{fig:figs7_architecture}
\end{figure*}

Fig.\ref{fig:figs7_architecture}(a) shows the overall flow of performing the genome sequencing with HDC, where the reference genome library are first encoded into random binary hypervectors and then stored into 1FeFET CAM array, as shown in Fig.\ref{fig:figs7_architecture}(c). Once a query genome comes in to check whether it exists in the reference library, it also goes through the hypervector encoder, and the resulting hypervector is applied as a search query to the CAM array. 
The encoder, shown in Fig.\ref{fig:figs7_architecture}(b), maps the genome sequence into almost orthogonal hypervectors such that only similar genome sequences to the query have small distances. 
By setting an appropriate distance threshold to the output of CAM array sense amplifier, whether the reference library contains the query genome can be quickly answered within a single CAM search. Fig.\ref{fig:figs7_architecture}(d) shows the evaluation results of our proposed associative memory architecture for HDC genome sequencing tasks. 
The results suggest that with the  associative memory implementation performing the distance threshold based approximate search, our proposed search engine can achieve on average 89.9x (71.9x) faster and 66.5x (30.7x) higher energy efficiency as compared to the state-of-the-art alignment tools NVBIO (GPU-BLAST) approach.

\end{document}